\documentclass[sigconf,screen]{acmart}

\usepackage[ruled,vlined]{algorithm2e}
\usepackage[group-separator={,}]{siunitx}
\include{pythonlisting}

\AtBeginDocument{%
  \providecommand\BibTeX{{%
    \normalfont B\kern-0.5em{\scshape i\kern-0.25em b}\kern-0.8em\TeX}}}

\usepackage{url}

\usepackage{breakurl}
\begin{document}

\title{SeismographAPI: Visualising Temporal-Spatial Crisis Data}

\author{Raphael Lepuschitz}
\authornote{Both authors contributed equally to this work.}
\email{raphael.lepuschitz@gmail.com}
\affiliation{%
  \institution{University of Innsbruck}
  \city{Innsbruck}
  \country{Austria}
}

\author{Niklas Stoehr}
\authornotemark[1]
\email{niklas.stoehr@inf.ethz.ch}
\affiliation{%
  \institution{ETH Zurich}
  \city{Zurich}
  \country{Switzerland}
}






\begin{abstract}
Effective decision-making for crisis mitigation increasingly relies on visualisation of large amounts of data. While interactive dashboards are more informative than static visualisations, their development is far more time-demanding and requires a range of technical and financial capabilities. There are few open-source libraries available, which is blocking contributions from low-resource environments and impeding rapid crisis responses. To address these limitations, we present \href{https://github.com/conflict-AI/seismographAPI}{\textit{SeismographAPI}}, an open-source library for visualising temporal-spatial crisis data on the country- and sub-country level in two use cases --- \href{https://conflict-ai.github.io/seismographAPI/conflict-map.html}{\textit{Conflict Monitoring Map}} and \href{https://conflict-ai.github.io/seismographAPI/covid-map.html}{\textit{Pandemic Monitoring Map}}. The library provides easy-to-use data connectors, broad functionality, clear documentation and run time-efficiency.
\end{abstract}



\maketitle

\section{Introduction}


For mitigating large-scale crises such as armed conflicts, pandemics and natural disasters, incorporation of data in decision-making is becoming indispensable \cite{beck_improving_2000,sornette_endogenous_2006,weidmann_predicting_2010,obrien_crisis_2010, falck_measuring_2020}. However, insights from large amounts of data remain untapped if they are not detected and communicated by means of intuitive, accurate and preferably interactive scientific visualisation \cite{piburn_world_2015, kim_explaining_2017}. Particularly, the development of interactive visualisation dashboards requires a broad skill set, ranging from statistical, design and programming knowledge to domain expertise \cite{lam_empirical_2012}. Academic environments, non-governmental and humanitarian aid organisations often lack the required resources which hinders urgently needed contributions. The demand for quick crisis responses stands in stark contrast to time-consuming, expensive development stages. \href{https://github.com/conflict-AI/seismographAPI}{\textit{SeismographAPI}} is an actively maintained, open-source library for the visualisation of temporal-spatial crisis data that combines plug-and-play visualisations with versatile functionality.


\begin{figure}[t]
    \centering
    \includegraphics[width=\columnwidth]{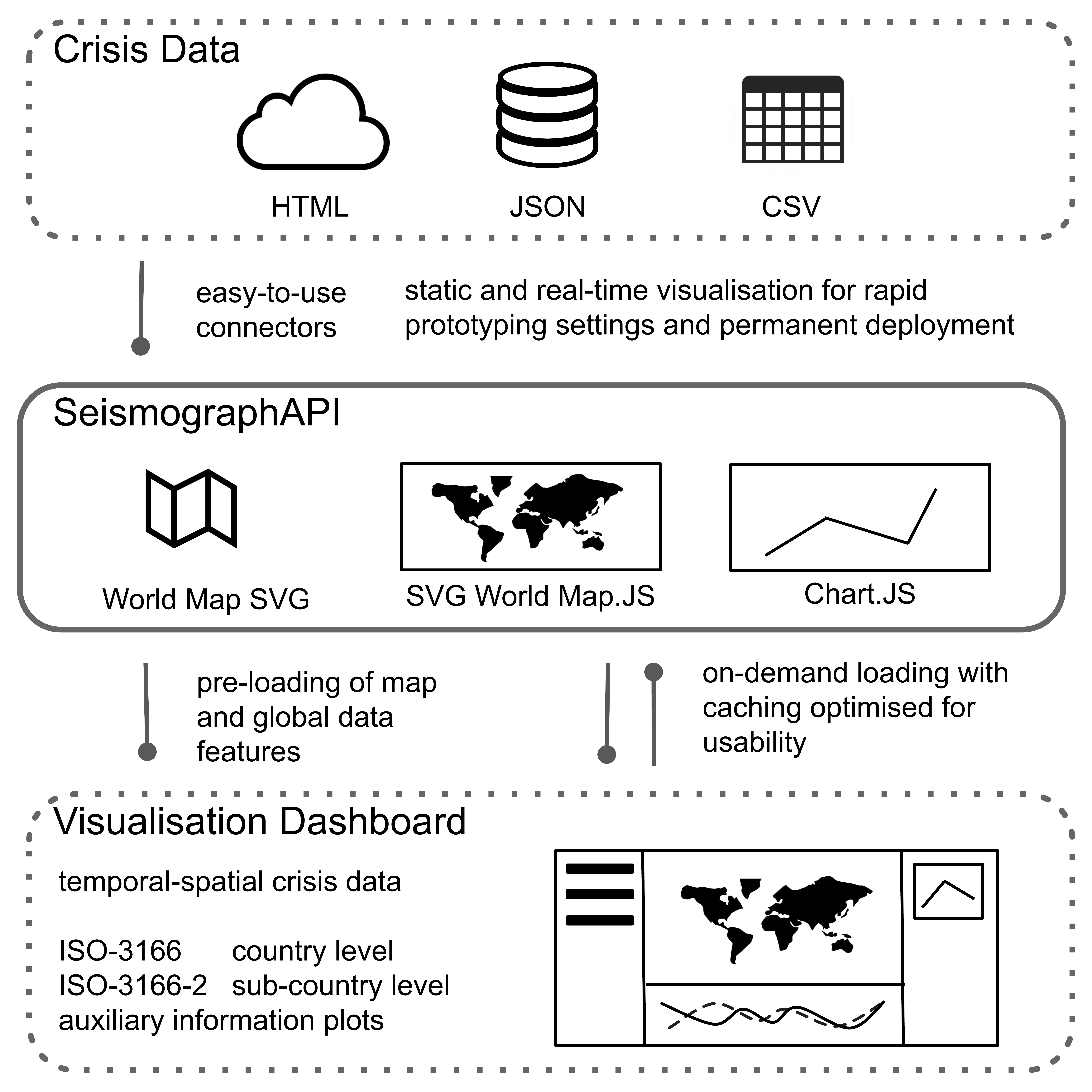}
    \caption{Technical overview of \href{https://github.com/conflict-AI/seismographAPI}{\textit{SeismographAPI}}}
    \label{fig:overview}
\end{figure}

\section{Exemplary Use Cases}

\href{https://github.com/conflict-AI/seismographAPI}{\textit{SeismographAPI}} is designed for data analysts to identify patterns in rapid prototyping. Due to its run time and memory-efficiency, it can also be deployed as a permanent visualisation tool for use by decision-makers. To motivate and demonstrate \href{https://github.com/conflict-AI/seismographAPI}{\textit{SeismographAPI}}, we sketch out two practical use cases that are inspired by real-world visualisation needs \cite{weidmann_predicting_2010,obrien_crisis_2010, hegre_predicting_2013, stephany_corisk-index_2020, dong_interactive_2020}. 

\paragraph{\textbf{Conflict Monitoring.}}

With the help of \href{https://github.com/conflict-AI/seismographAPI}{\textit{SeismographAPI}}, we visualise a huge dataset comprising 20 years of conflict data on 141 countries, constructed from \href{https://acleddata.com/#/dashboard}{\textit{ACLED}} \cite{raleigh_introducing_2010} and \href{https://ucdp.uu.se}{\textit{UCDP GED}} \cite{sundberg_introducing_2013} data. Per country and month, our dataset features 60 socio-economic and political indicators, which are all displayed in our \href{https://conflict-ai.github.io/seismographAPI/conflict-map.html}{\textit{Conflict Monitoring Map}}.

\paragraph{\textbf{Pandemic Monitoring}}

Our second demonstration case is the \href{https://conflict-ai.github.io/seismographAPI/covid-map.html}{\textit{Pandemic Monitoring Map}}, a visualisation of \textit{COVID-19} infection numbers. The \href{https://github.com/CSSEGISandData/COVID-19}{data} is borrowed from Johns Hopkins University \cite{dong_interactive_2020}.

\begin{figure*}[t]
    \centering
    \includegraphics[width=\textwidth]{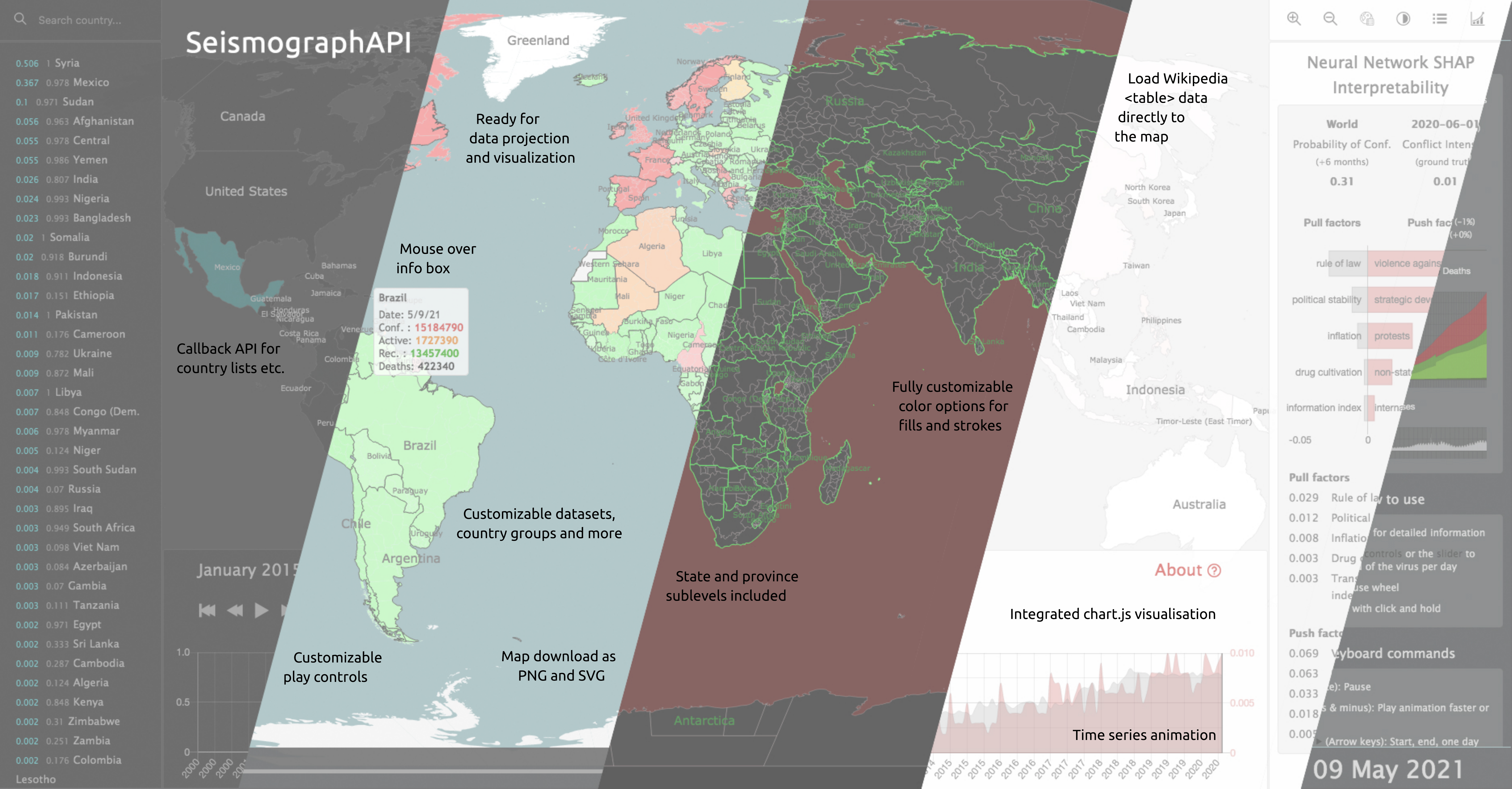}
    \caption{\href{https://conflict-ai.github.io/seismographAPI/conflict-map.html}{\textit{Conflict Monitoring Map}} and \href{https://conflict-ai.github.io/seismographAPI/covid-map.html}{\textit{Pandemic Monitoring Map}}: two exemplary use cases of the \href{https://github.com/conflict-AI/seismographAPI}{\textit{SeismographAPI}}}
    \label{fig:demo}
\end{figure*}

\section{Main Functionality}

\paragraph{\textbf{World Map (center).}}

The SVG Choropleth map represents the core part of \href{https://github.com/conflict-AI/seismographAPI}{\textit{SeismographAPI}}. It allows visualising data at the country- and subcountry-level (political subdivisions) based on the \href{https://www.iso.org/iso-3166-country-codes.html}{ISO-3166} and \href{https://www.iso.org/iso-3166-country-codes.html}{ISO-3166-2} norm. Additional information, such as country-level infection numbers, can be easily displayed on click and hover as exemplified in the \href{https://conflict-ai.github.io/seismographAPI/covid-map.html}{\textit{Pandemic Monitoring Map}}.

\paragraph{\textbf{Time Series Chart (bottom).}}

The time series chart not only visualises, but allows navigating the temporal dimension. For instance, the \href{https://conflict-ai.github.io/seismographAPI/conflict-map.html}{\textit{Conflict Monitoring Map}} even features two time lines, one showing the prediction and another showing the ground truth conflict intensity. When hovering or clicking a point in time, all other panels synchronise. With the help of the ``play'' controls, users can watch all data panels as they change over time in a time-machine manner.

\paragraph{\textbf{Auxiliary Information Panel (right).}}

At the top of the auxiliary information panel, our library provides a menu allowing to interactively customise the dashboard. Users can hide information and panels, such as country names and the country list on the left hand side, zoom-in, choose a night mode and open a ``help" window. To simplify the interface between analysis, report and decision-making, the library has built-in functionality for screen recording. Due to tight integration with \href{https://www.chartjs.org/}{Chart.js}, any chart visualisation can be selected and displayed in the right-hand panel based on data suitability and information needs. For instance, the \href{https://conflict-ai.github.io/seismographAPI/conflict-map.html}{\textit{Conflict Monitoring Map}} displays the most important data features considered for conflict prediction as a horizontal bar chart. The \href{https://conflict-ai.github.io/seismographAPI/covid-map.html}{\textit{Pandemic Monitoring Map}} relies on stacked line charts to map out infection numbers.

\section{Technical Background}

\paragraph{\textbf{Run time and Memory.}}

\href{https://github.com/conflict-AI/seismographAPI}{\textit{SeismographAPI}} builds upon two fast, open-source libraries, \href{https://www.chartjs.org/}{Chart.js} and \href{https://github.com/raphaellepuschitz/SVG-World-Map}{SVG World Map JS}. The time required for data loading is mainly determined by the size of the central SVG world map: \textasciitilde1,3 MB for {ISO-3166-2} country-level and \textasciitilde3,8 MB including all subdivision data. Depending on the chosen map, rendering starts between 300ms and 800ms, document completion is done between 400ms and 2.6s and the full loading time varies from \textasciitilde3s to \textasciitilde10s. To optimise loading and usability, \href{https://github.com/conflict-AI/seismographAPI}{\textit{SeismographAPI}} can also be initialised asynchronously with the JavaScript \href{https://developer.mozilla.org/en-US/docs/Web/JavaScript/Reference/Operators/await}{async/await/resolve} method. After the first initialisation of the map, this enables loading data chunks on demand, which increases smoothness. This is demonstrated in the \href{https://conflict-ai.github.io/seismographAPI/conflict-map.html}{\textit{Conflict Monitoring Map}}, where all global conflict data (\textasciitilde1,1MB) is loaded at startup, but the large amount of detailed conflict data (\textasciitilde80KB per country, \textasciitilde21MB in total) is loaded asynchronously on request. Thus, \href{https://github.com/conflict-AI/seismographAPI}{\textit{SeismographAPI}} is able to visualise more than $N = \num{170000}$ data points in the \href{https://conflict-ai.github.io/seismographAPI/conflict-map.html}{\textit{Conflict Monitoring Map}} in \href{https://www.webpagetest.org/result/210509_AiDc81_73869360943cdc9a6a40f9dc250a20b8/}{about 3 seconds} or nearly $N = \num{400000}$ data points in the \href{https://conflict-ai.github.io/seismographAPI/covid-map.html}{\textit{Pandemic Monitoring Map}} in \href{https://www.webpagetest.org/result/210509_BiDc1S_4760793a346f09615cae9fa5ac5124e6/}{about 10 seconds}.

\paragraph{\textbf{Ease of Use.}}

With an intuitive interface and simple data connectors, \href{https://github.com/conflict-AI/seismographAPI}{\textit{SeismographAPI}} is designed for ease of use in common visualisation tasks and workflows. Data can be loaded directly via JSON, CSV or as an HTML table. We even offer a \href{https://pandas.pydata.org}{Pandas} extension to load \href{https://pandas.pydata.org/docs/reference/api/pandas.DataFrame.html}{Pandas Dataframes} (as JSON) and \href{https://en.wikipedia.org/wiki/Help:Table}{Wikipedia tables}. The library features clear readme instructions and rich documentation. 

\section{Conclusion}

Future versions will include more data connectors, default charts, more detailed guidelines for deployment and options for switching between different data within one map. We presented \href{https://github.com/conflict-AI/seismographAPI}{\textit{SeismographAPI}}, an open-source library aimed at reducing resource constraints and easing swift data visualisation, thereby improving data-driven decision-making for humanitarian purposes.

\bibliographystyle{ACM-Reference-Format}
\bibliography{reference}

\appendix
\section{Appendix}
\subsection{Links}

\textit{SeismographAPI Github}\\ \href{https://github.com/conflict-AI/seismographAPI}{https://github.com/conflict-AI/seismographAPI}\\
\textit{Conflict Monitoring Map}\\ \href{https://conflict-ai.github.io/seismographAPI/conflict-map.html}{https://conflict-ai.github.io/seismographAPI/conflict-map.html}\\
\textit{Pandemic Monitoring Map}\\ \href{https://conflict-ai.github.io/seismographAPI/covid-map.html}{https://conflict-ai.github.io/seismographAPI/covid-map.html}\\

\end{document}